\documentclass[twocolumn]{aastex63}

\usepackage{amsmath}
\usepackage{chngcntr}
\usepackage{stix}
\usepackage{enumitem}
\usepackage{xspace}

\newcommand{\halpha}{H\ensuremath{\alpha}\xspace}
\newcommand{\angstrom}{\text{\normalfont\AA}}

\newcommand{\Lacc}{\ensuremath{L_\mathrm{acc}}\xspace}

\newcommand{\Lhalpha}{\ensuremath{L_\mathrm{H\alpha}}\xspace}

\newcommand{\Lsol}{\ensuremath{L_{\odot}}\xspace}
\newcommand{\abaur}{AB~Aur\xspace}

\graphicspath{{./}{figures/}{Figures/}}

\received{}
\revised{}
\accepted{}
\shorttitle{High-contrast Imaging of AB Aur in \halpha}
\shortauthors{Zhou, Sanghi et al.}

\begin{document}
\title{HST/WFC3 H$\alpha$ Direct-Imaging Detection of a Point-like Source in the Disk Cavity of AB Aur}

\correspondingauthor{Yifan Zhou}
\email{yifan.zhou@utexas.edu}
\author[0000-0003-2969-6040]{Yifan Zhou}
\altaffiliation{These authors contributed equally to this work.}
\affiliation{Department of Astronomy, The University of Texas at Austin,  2515 Speedway, Stop C1400 Austin, TX 78712, USA}
\affiliation{51 Pegasi b Fellow}
\author[0000-0002-1838-4757]{Aniket Sanghi}
\altaffiliation{These authors contributed equally to this work.}
\affiliation{Department of Astronomy, The University of Texas at Austin,  2515 Speedway, Stop C1400 Austin, TX 78712, USA}
\author[0000-0003-2649-2288]{Brendan P. Bowler}
\affiliation{Department of Astronomy, The University of Texas at Austin,  2515 Speedway, Stop C1400 Austin, TX 78712, USA}

\author[0000-0002-4392-1446]{Ya-Lin Wu}
\affiliation{Department of Physics, National Taiwan Normal University, Taipei 116, Taiwan}
\affiliation{Center of Astronomy and Gravitation, National Taiwan Normal University, Taipei 116, Taiwan}

\author[0000-0002-2167-8246]{Laird M. Close}
\affiliation{Department of Astronomy, University of Arizona, Tucson, AZ 85719, USA}

\author[0000-0002-7607-719X]{Feng Long}
\affiliation{Center for Astrophysics \textbar\, Harvard \& Smithsonian, 60 Garden St., Cambridge, MA 02138, USA}

\author[0000-0002-4479-8291]{Kimberly Ward-Duong}
\affiliation{Department of Astronomy, Smith College, Northampton, MA 01063, USA}
\author[0000-0003-3616-6822]{Zhaohuan Zhu}
\author[0000-0001-9811-568X]{Adam L. Kraus}
\affiliation{Department of Astronomy, The University of Texas at Austin,  2515 Speedway, Stop C1400 Austin, TX 78712, USA}
\author[0000-0002-7821-0695]{Katherine B. Follette}
\affiliation{Department of Physics and Astronomy, Amherst College, Amherst, MA 01003, USA}
\author[0000-0001-7258-770X]{Jaehan Bae}
\affiliation{Department of Astronomy, University of Florida, Gainesville, FL 32611, USA}

\begin{abstract}
  Accreting protoplanets enable the direct characterization of planet formation. As part of a high-contrast imaging search for accreting planets with the Hubble Space Telescope (HST) Wide Field Camera 3, we present H$\alpha$ images of AB Aurigae (AB Aur), a Herbig Ae/Be star harboring a transition disk. The data were collected in two epochs of direct-imaging observations using the F656N narrow-band filter. After subtracting the point spread function of the primary star, we identify a point-like source located at a P.A. of $182.5^{\circ}\pm1.4^{\circ}$ and a separation of $600\pm22$~mas relative to the host star.  The position is consistent with the recently identified protoplanet candidate AB Aur b. The source is visible in two individual epochs separated by ${\sim}50$ days and the H$\alpha$ intensities in the two epochs agree. The H$\alpha$ flux density is $F_{\nu}=1.5\pm0.4$~mJy, $3.2\pm0.9$ times of the optical continuum determined by published HST/STIS photometry. In comparison to PDS 70 b and c, the H$\alpha$ excess emission is weak. The central star is accreting and the stellar H$\alpha$ emission has a similar line-to-continuum ratio as seen in AB Aur b.  We conclude that both planetary accretion and scattered stellar light are possible sources of the H$\alpha$ emission, and the H$\alpha$ detection alone does not validate AB Aur b as an accreting protoplanet.  Disentangling the origin of the emission will be crucial for probing planet formation in the AB Aur disk.
\end{abstract}

  \section{Introduction}

  Two of the most fundamental goals of planet formation studies are to understand how and when planets gain mass. The planetary mass assembly process can be directly constrained by characterizing actively accreting protoplanets. So far, the discoveries of protoplanets are limited to PDS~70 b and c, two gas giants directly imaged within the same planetary system \citep{Keppler2018,Haffert2019}. A plethora of follow-up studies on this system confirmed the presence of circumplanetary disks \citep{Isella2019,Benisty2021}, estimated the mass accretion rates \citep[e.g.,][]{Wagner2018, Haffert2019, Hashimoto2020,Zhou2021}, investigated the planet-disk interactions \citep[e.g.,][]{Bae2019}, and constrained the planetary orbits \citep{Wang2021}. Despite this steady progress, these studies are confined to one protoplanetary system and only represent a single outcome of the planet formation process observed at an instant in time.

  To expand the sample of accreting planets, we launched the Hubble Accreting Luminous Protoplanets in H-Alpha (HALPHA) Survey (Program ID: 16651\footnote{Program information can be found at \url{https://www.stsci.edu/cgi-bin/get-proposal-info?id=16651&observatory=HST}}, PI: Zhou) to search for accreting planets in ten transition disk systems. We are exploiting the recently demonstrated optical and ultraviolet (UV) high-contrast imaging capibility of Wide Field Camera 3 (WFC3) UVIS \citep{Zhou2021,Sanghi2022} and using the narrow-band F656N (\halpha) filter to look for \halpha-emitting planets. Accretion onto planets produces strong hydrogen line emission, reduces planet-star contrasts, and can in principle improve search efficiency. Any detections result in \halpha luminosities (\Lhalpha) and accretion rate estimates of candidate planets.

  The strategy of observing transition disk systems is motivated by models showing that the disk gaps and cavities are consistent with being sculpted by multiple giant planets \citep{Dodson2011,Close2020}. These models are broadly supported by the discoveries and characteristics of PDS 70 b and c \citep{Bae2019}. We identified ten systems that harbor giant cavities demonstrated in ALMA dust continuum images \citep{Francis2020}. The host stars are generally faint or located in the northern hemisphere and, consequently, are not easily accessible by ground-based visible-light adaptive-optics (AO) systems.


  \begin{figure*}[!th]
    \centering
    \includegraphics[width=\textwidth]{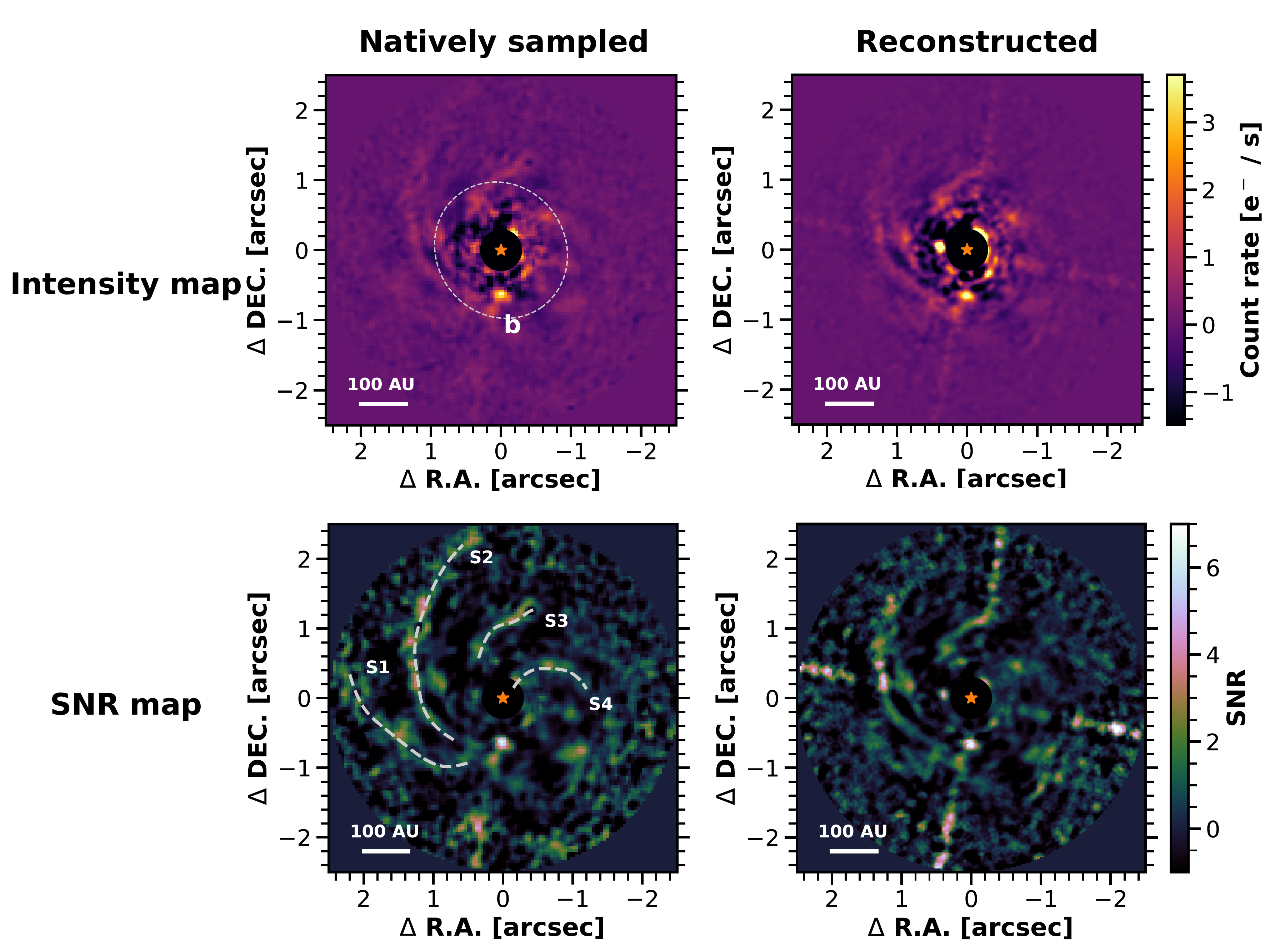}
    \caption{Primary-subtracted images (upper) and the SNR maps (lower) of AB Aur observed by HST/WFC3/UVIS in the F656N filter. The left column shows the primary-subtracted natively sampled images (pixel scale\,=\,40~mas) and the right column shows the  reconstructed images (pixel scale\,=\,20~mas). The upper panels show the intensity maps (in unit of $\mathrm{e^{-}s^{-1}}$). These images, as well as the upper panels of Figure~\ref{fig:epoch}, the left panel of Figure~\ref{fig:injection}, and Figure~\ref{fig:detection} are smoothed using a Gaussian kernel with $\mbox{FWHM}=0.07''$ ($1.0\times$ PSF FWHM). Smoothing improves visualization by reducing high-frequency noise but is not applied in the sensitivity or contrast curve estimates. The lower panels are the SNR maps. For all four images, the color maps have a linear stretch. The central $r=0.3''$ radius circle is masked out, because in this region the detection sensitivity is poor due to high-contrasts and saturation. In the upper-left panel, an ellipse marks the transition disk cavity identified by \citet{Francis2020}. In the lower-left panel, gray dashed lines indicate the spiral disk structures (S1 to S4) detected in our images. The point-like source at $0.6''$ south of AB Aur is visible in all panels. }
    \label{fig:image}
  \end{figure*}

  \begin{figure*}[!th]
    \centering
    \includegraphics[width=\textwidth]{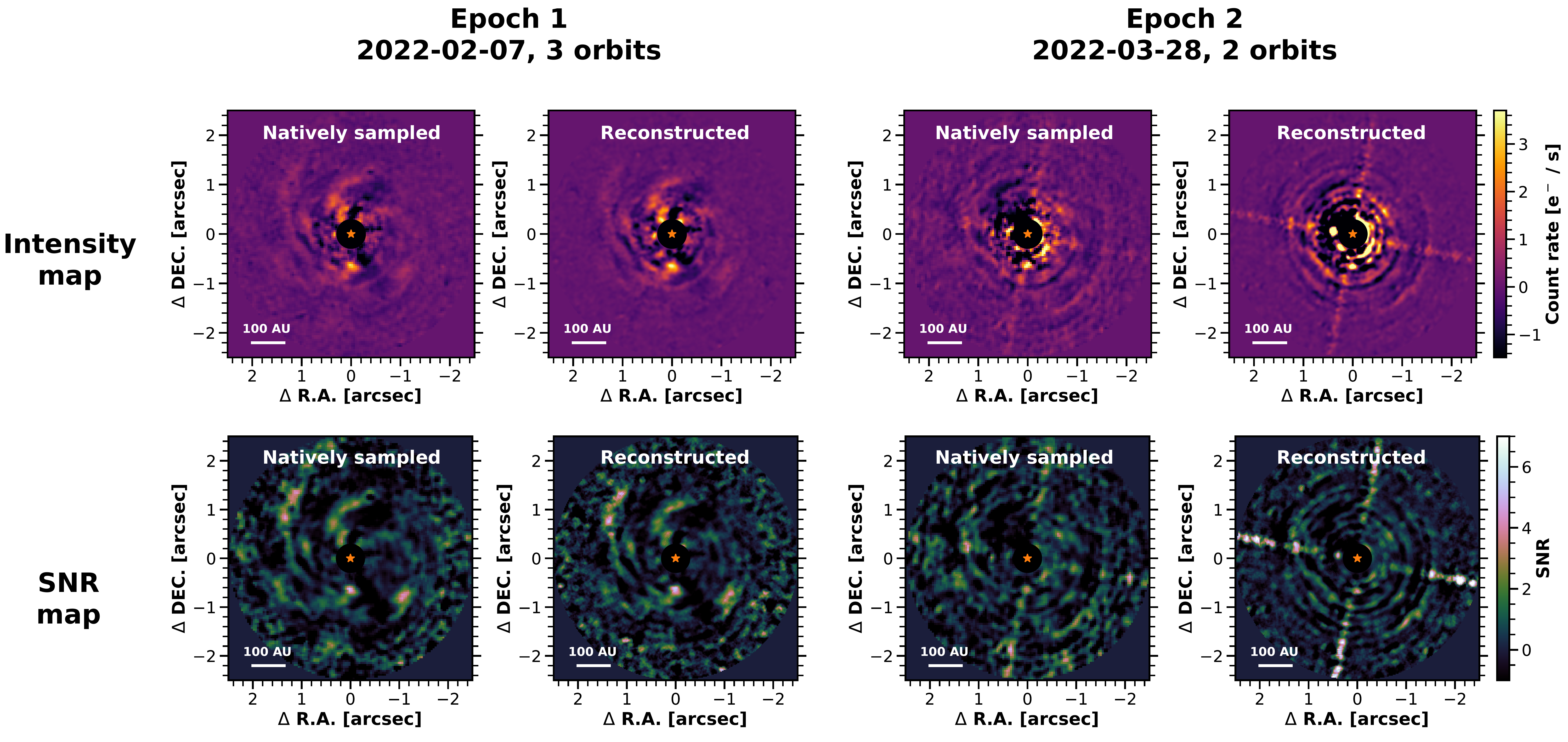}
    \caption{Primary-subtracted images and SNR maps in Epochs 1 (left four panels) and 2 (right four panels). The intensity and SNR maps obtained from processing the natively sampled and reconstructed images are organized in the same way as Figure~\ref{fig:image}. The point source is consistently recovered in both epochs, although the detection SNRs are below 5 in the second epoch due to shorter integration time. Systematic residuals near the diffraction spikes are also prominent in the second epoch.}
    \label{fig:epoch}
    \end{figure*}

  The first target of the survey is AB Aurigae (AB Aur), a  Herbig Ae/Be star \citep{DeWarf2003} with a spectral type of A0, a mass of $2.4\pm0.2M_{\odot}$, and an isochronal age of $6.0^{+2.5}_{-1.0}$~Myr\footnote{The age is estimated with the PARSEC isochrones \citep{Bressan2012} and the observationally constrained $L_{\mathrm{bol}}$ and spectral type of AB Aur \citep{Herczeg2014}.}. It is actively accreting at a rate of $\dot{M}\approx10^{-7}M_{\odot}\mathrm{yr^{-1}}$ \citep{Garcia-Lopez2006} and exhibits significant variability (${>}10\%$) in both broadband photometry and the \halpha line \citep{Harrington2007,Cody2013}. AB Aur hosts a transition disk that includes a highly structured inner disk with multiple spirals \citep[e.g.,][]{Boccaletti2020,Jorquera2022} and a cavity with an outer edge at 156~AU \citep{Francis2020}. Several spiral structures have been identified in the inner disk and their presence has been attributed to ongoing planet formation \citep{Fukagawa2004, Oppenheimer2008, Boccaletti2020}. Specifically, \citet{Tang2017} identified two spirals in high-resolution ALMA  $^{12}$CO $J{=}2{-}1$ emission observations and suggested that the spiral morphology could be explained by tidal disturbance caused by a companion located at 60 to 80 AU ($0.4''$ to $0.6''$ angular separation) and a ${\sim}180^{\circ}$ position angle (P.A.). Recently, \citet{Currie2022} presented evidence for an embedded protoplanet, referred to as AB Aur b, in visible and infrared images near the location predicted by \citet{Tang2017}. In this \emph{Letter}, we report results from two epochs of HST/WFC3 F656N (\halpha) high-contrast imaging observations of AB Aur.

  \section{Observations}

  \subsection{HST Observations}
  \abaur was observed by HST/WFC3/UVIS on UT 2022-02-07 (Epoch 1) and 2022-03-28 (Epoch 2) for three (Orbits 1 to 3) and two orbits (Orbits 4 and 5), respectively. A 0.5-pixel (20~mas), four-point-box dithering strategy was adopted in all five orbits to enable the reconstruction of Nyquist-sampled images. In every orbit, HST sequentially pointed at each position and took eleven identical 2.7~s exposures in the F656N filter (\halpha narrow-band; $\lambda_{\mathrm{c}}=6561.5\,\angstrom$, $\Delta\lambda=17.9\,\angstrom$). The exposures were captured by the \texttt{c512c} subarray, which has a native pixel scale of 40~mas and a field of view of $20''\times20''$.  Due to guide star acquisition failures, the first frame of Epoch 1 and the first two frames of Epoch 2 did not acquire the target. We discarded these images in our analysis. In total, the HST observations comprised 217 raw frames, amounting to 586~s of on-source exposure time.

  HST's roll angles (the V3 axis orientation) were $246.3^{\circ}$, $271.3^{\circ}$, $246.3^{\circ}$, $260.6^{\circ}$, and $260.6^{\circ}$  in Orbits 1 to 5, respectively. The angular differentials between any two distinct roll angles are $10.7^{\circ}$, $14.3^{\circ}$, and $25^{\circ}$. At a separation of $0.6''$, these correspond to spatial displacements of 108, 149, and 262~mas, or 1.6, 2.2, and 3.7 times the full width at half maximum (FWHM) of the F656N point spread function (PSF), respectively. These large azimuthal displacements enable the use of angular differential imaging (ADI, e.g., \citealt{Marois2006}) to subtract the PSF of the host star between any pair of images with different telescope rolls.

  The 2022-03-28 observations were a repeat of the first and third orbits of the 2022-02-07 observations that suffered from a fine-guiding sensor failure resulting in 0.1 to 0.3 pixel pointing errors. This issue impairs image reconstruction accuracy at small scales (${<}15\,\mbox{mas}$) but is not detrimental to this study. Therefore, we present results from all five orbits.

  \subsection{High-resolution Spectroscopy of AB Aur}

  The HST images do not constrain the continuum or resolve the \halpha line. To precisely characterize the \halpha emission from the host star, we obtained an optical \'echelle spectrum of AB Aur using the Tull Coud\'e spectrograph \citep{Tull1995} on the 2.7-m Harlan J. Smith Telescope at McDonald Observatory on UT 2019-09-26. The exposure time was 600~s and conducted with a $1.2''$ slit, resulting in a resolving power of $R=60,000$ and a wavelength span from $3870\,\angstrom$ to $10,450\,\angstrom$ in a total of 56 orders. The wavelength calibration was carried out with a ThAr emission lamp spectrum.

  The Tull spectrum is reduced with a custom pipeline. We adopt the order that spans from $6547.7\,\angstrom$ to $6656.7\,\angstrom$ and contains the \halpha line. The spectrum is normalized by the continuum determined by the best-fitting fifth order polynomial of the line-excluded region ($\lambda<6550\angstrom$ or $\lambda>6580\angstrom$). We compute the synthetic photometry in the F656N filter using the \texttt{pysynphot} package and find that the band-average flux is 2.58 times of the continuum.

  \section{Data Reduction}

  \begin{figure*}
    \centering
    \includegraphics[width=\textwidth]{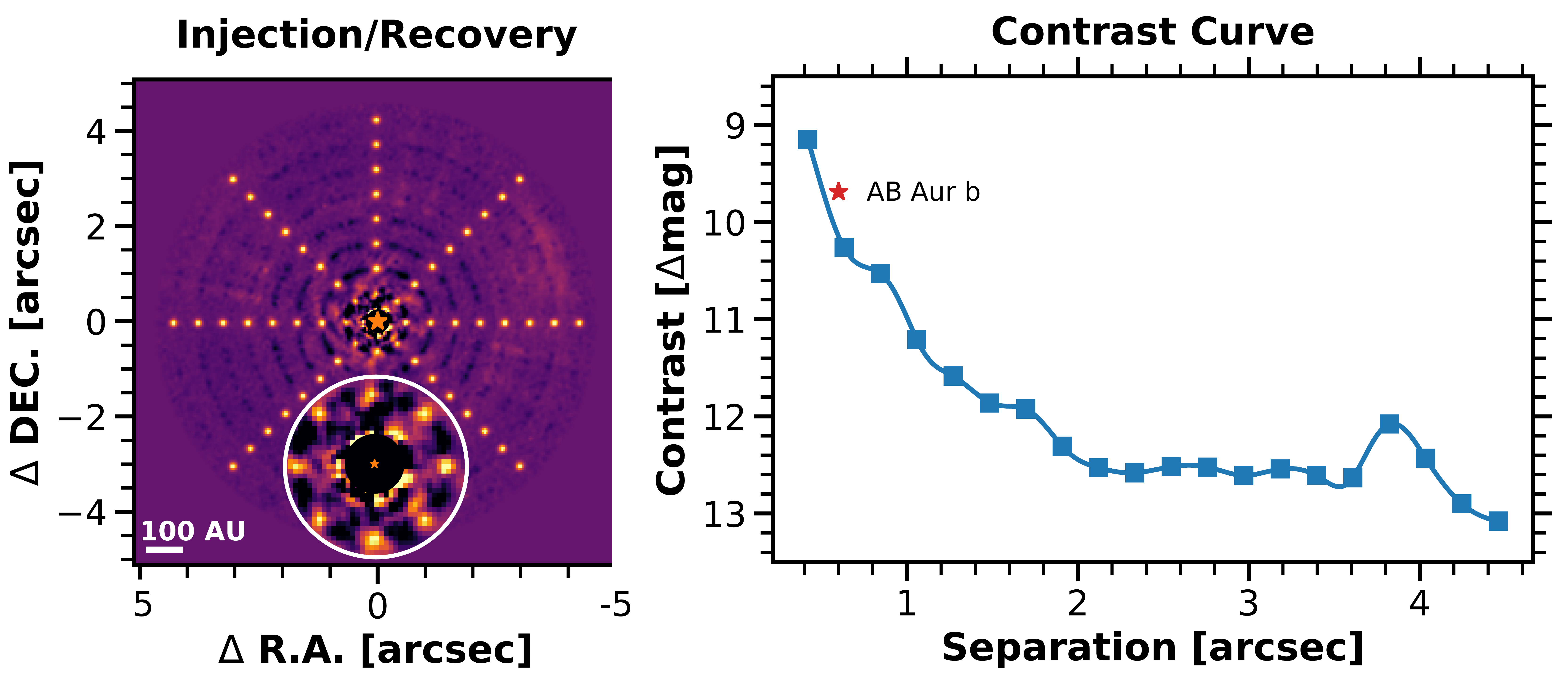}
    \caption{The injection-and-recovery image (left) and the $5\sigma$ contrast curve (right).
      The injection-and-recovery test is conducted on the natively sampled images. Injected artificial planets with the same brightness as the one measured in the point-like source are consistently recovered and their recovered photometry is used in correcting companion flux loss in primary subtraction and calibrating the contrast curve. The inset in the left panel is the inner $r=0.7''$ circular region of the injection-and-recovery image, where the artificial sources have the same angular separations as AB Aur b.
      In the right panel, the $5\sigma$ contrast measurements are shown in squares, which are connected by a spline interpolation (solid line).
      The loss of sensitivity at $3.6''$ to $4.0''$ is due to a known filter ghost.}
    \label{fig:injection}
  \end{figure*}

  Our data reduction starts with the \texttt{flc} files downloaded from the HST archive. The nominal bias, dark, flat field calibrations, as well as charge transfer efficiency correction are performed by the \texttt{CalWFC3} pipeline. We first identify cosmic-ray (CR) affected pixels using the data quality flag (DQ=1024) and replace these pixels with two-dimensional linear interpolations of the respective neighboring pixels. Every set of four-point dithered and CR-corrected frames are processed by a Fourier-interlacing pipeline \citep{Lauer1999a,Zhou2021} to construct one $2\times$ finer sampled image. This pipeline is optimized for point-source PSFs and distinct from the drizzling method that is part of the regular data reduction performed at STScI.

  We then independently perform a complete set of primary subtraction procedures on the natively sampled (pixel scale${=}40$~mas) and reconstructed (pixel scale${=}20$~mas) image sets. First, the image cube is registered by the primary star centroids determined by the \texttt{photutils} package’s \texttt{centroid\_2dg} function \citep{Robitaille2013}. Then, the images are split into two annuli that have inner/outer radii of $0.3''/0.7''$ and $0.7''/2.5''$, respectively. We subtract the host star PSF in each annulus using the Karhunen-Lo\`eve Image Projection (KLIP, \citealt{Soummer2012}) algorithm with a custom pipeline. For each image, all frames obtained at a different roll angle are used as references and the number of KLIP components is equal to the number of reference frames. We have also experimented with limiting the KLIP components to 5, 10, and 15 and find that the variation has negligible effect on the subtraction results. After primary subtraction, we rotate all frames to align their $y$-axes to the true north and co-add the aligned frames to form the final image. The co-adding step is also performed for the two epochs individually. In total, we create six primary-subtracted images for two pixel scales each with three sets of co-adds (the combined set: Figure~\ref{fig:image}; individual epochs: Figure~\ref{fig:epoch}).

  We convert the primary-subtracted images into signal-to-noise ratio (SNR) maps for point-source detections. The SNR is defined as the ratio between the flux integrated in a 1~FWHM radius aperture and the standard deviation of fluxes along the remaining azimuthal region sampled by non-overlapping apertures. Biases induced by small-number statistics were corrected based on \citet{Mawet2014}. The SNR maps are shown in Figures~\ref{fig:image} and \ref{fig:epoch} under the respective primary-subtracted images. Substantial residuals appear near the diffraction spikes in the reconstructed images, which are related to the telescope pointing errors. 

  A point-like source emerges in every PSF-subtracted image at an identical location ${\sim}600$~mas south of the primary star. We apply the KLIP forward modeling method \citep{Pueyo2016} on the natively sampled images to precisely determine the astrometry and photometry of this source and correct measurement biases introduced by primary subtraction. Synthetic WFC3 UVIS2 F656N PSFs are generated with the TinyTim package \citep{Krist2011} as the point-source model. We inject a negative-flux PSF to the original frames and optimize the P.A., separation, and flux of the injection so that the residual sum of squares in a 3-pixel radius circular aperture minimizes. 


    To estimate the astrometric and photometric uncertainties, as well as the KLIP throughput, we conduct injection-and-recovery tests on the natively sampled images. Positive-flux TinyTim PSFs are injected at P.A.s of $0^{\circ}$, $\pm45^{\circ}$, $\pm90^{\circ}$, and $\pm135^{\circ}$ spanning eight equal-interval separations between $0.62''$ and $4.26''$ (Figure~\ref{fig:injection}). These PSFs have the same flux as the point-like source. We subtract the primary PSF  and then measure the positions and fluxes of the injected sources. At a specified separation, the average ratio between the injected and recovered fluxes is adopted as the throughput, and the standard deviations of the recovered fluxes, P.A.s, and separations are adopted to be the photometric and astrometric uncertainties. 

    Figure~\ref{fig:injection} includes the $5\sigma$ contrast curve that characterizes our detection sensitivity at angular separations spanning $0.4''$ to $4.5''$. In deriving the contrast curve,  we calculate the mean and standard deviation of the fluxes in non-overlapping 1-FWHM-radius apertures centered at a given separation. At 600~mas, the resolution element containing the detected point source is excluded. We follow the SNR definition in \citet{Mawet2014} to convert the mean and standard deviation into a flux corresponding to $\mbox{SNR}=5$ and normalize this value by the flux of AB Aur.  Finally, we divide the contrast by the KLIP throughput determined by positive PSF injections (Figure~\ref{fig:injection}) to correct for flux loss in primary subtraction.

  \section{Results}

\begin{figure}[!t]
    \centering
    \includegraphics[trim=1cm 0cm .5cm 0cm,clip,width=\columnwidth]{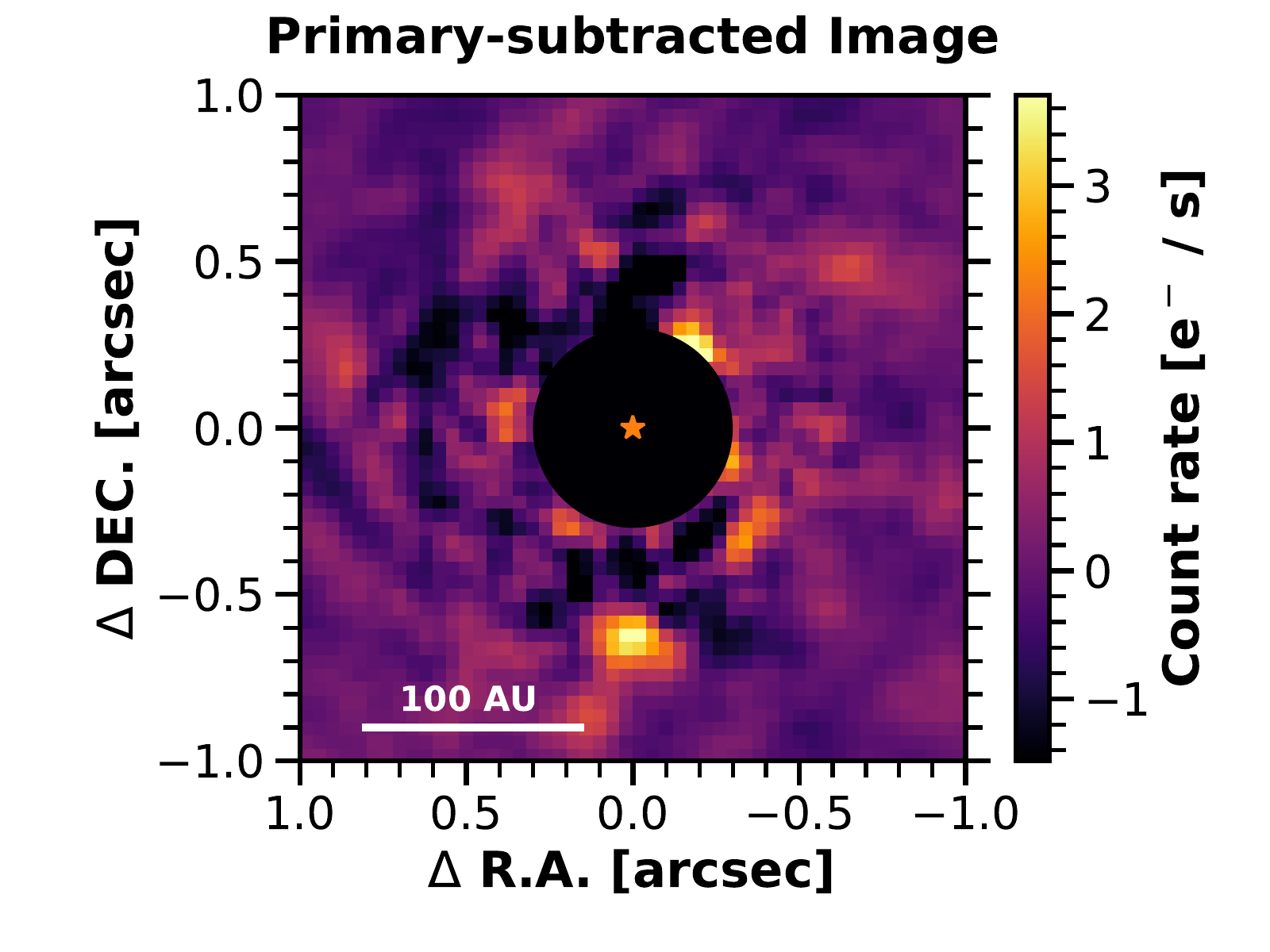}
    \caption{A zoomed-in ($2''\times2''$) natively sampled primary-subtracted image highlighting the detection of the point-like source. The source is at a $\mbox{P.A.}$ of $182^\circ$ and a separation of $600$~mas relative to the host star. This detection is at the location of AB Aur b, a candidate embedded planet recently reported by \citet{Currie2022}.}
    \label{fig:detection}
  \end{figure}

  We detect a point-like source near AB Aur in the primary-subtracted image (Figure~\ref{fig:detection}). In the natively sampled frames, the detection SNRs are 5.4, 3.8, and 6.4 in Epoch 1, Epoch 2, and the combined image, respectively; in the reconstructed images, the SNRs are 5.6, 4.4, and 7.8. The consistent results in natively sampled and reconstructed images, as well as in individual epochs, further strengthen the confidence in the detection. The source is located at a separation of $600\pm22$~mas ($93\pm3$~AU in projected physical distance) away from AB~Aur with a P.A. of $182.5^{\circ}\pm1.4^{\circ}$. This position is consistent with the protoplanet candidate AB Aur b reported in \citet{Currie2022}. Photometry on the combined image yields $64\pm16$ $\mathrm{e}^{-}$/s, corresponding to a band-averaged flux density of $F_{\nu}=1.5\pm0.4$~mJy or $F_{\lambda}=1.0\pm0.3\times10^{-15}$~erg/s/cm$^{2}$/$\angstrom$. Fitting a 2D Gaussian to the source yields a FWHM of 150~mas in the tangential direction and 127~mas in the radial direction, which are 2.1 and 1.8 times the FWHM of a F656N PSF. The FWHMs of recovered artificial companions are between 85 to 120 mas. The larger size of the detected source suggests it may trace slightly extended emission.
  
  The \halpha emission from the companion is consistent between the two epochs separated by 50 days, but our constraint on variability is weak. Photometric results of the individual epochs are $55\pm18$ $\mathrm{e}^{-}$/s and $76\pm30$ $\mathrm{e}^{-}$/s. The corresponding flux densities are $F_{\nu,1}=1.3\pm0.4$~mJy and $F_{\nu,2}=1.8\pm0.7$~mJy or $F_{\lambda,1}=0.9\pm0.3\times10^{-15}$~erg/s/cm$^{2}$/$\angstrom$ and $F_{\lambda,2}=1.2\pm0.5\times10^{-15}$~erg/s/cm$^{2}$/$\angstrom$. Based on the joint flux uncertainty between the two epochs, we reject variability with amplitude greater than 50\%. Using the same HST images, we precisely determine the \halpha variability of the star AB~Aur to be $13.1\pm0.2\%$, consistent with the known variability of the host star \citep{Harrington2007,Cody2013}.

  Four spiral-like scattered-light features are detected at $\mathrm{SNR}$ values between 2 to 3. In Figure~\ref{fig:image}, we mark these structures as S1 to S4. These spiral features have all been previously identified in high-contrast imaging observations of AB Aur \citep[e.g.,][]{Fukagawa2004, Boccaletti2020}. Notably, the leading knots of S1 and S2 point towards the position of the point-like source.

  We do not detect any sources at $2.75''$ or $3.72''$ reported in \citet{Currie2022}. The $5\sigma$ upper limits in our combined WFC3 frames are $0.098$~mJy and $0.091$~mJy, respectively, and are 4.8 and 22 times the flux measured by HST/STIS \citep{Currie2022}. Therefore, our observations do not have sufficient sensitivity to determine the nature of the two sources.

  \begin{figure*} [!ht]
    \centering
    \includegraphics[width=\columnwidth]{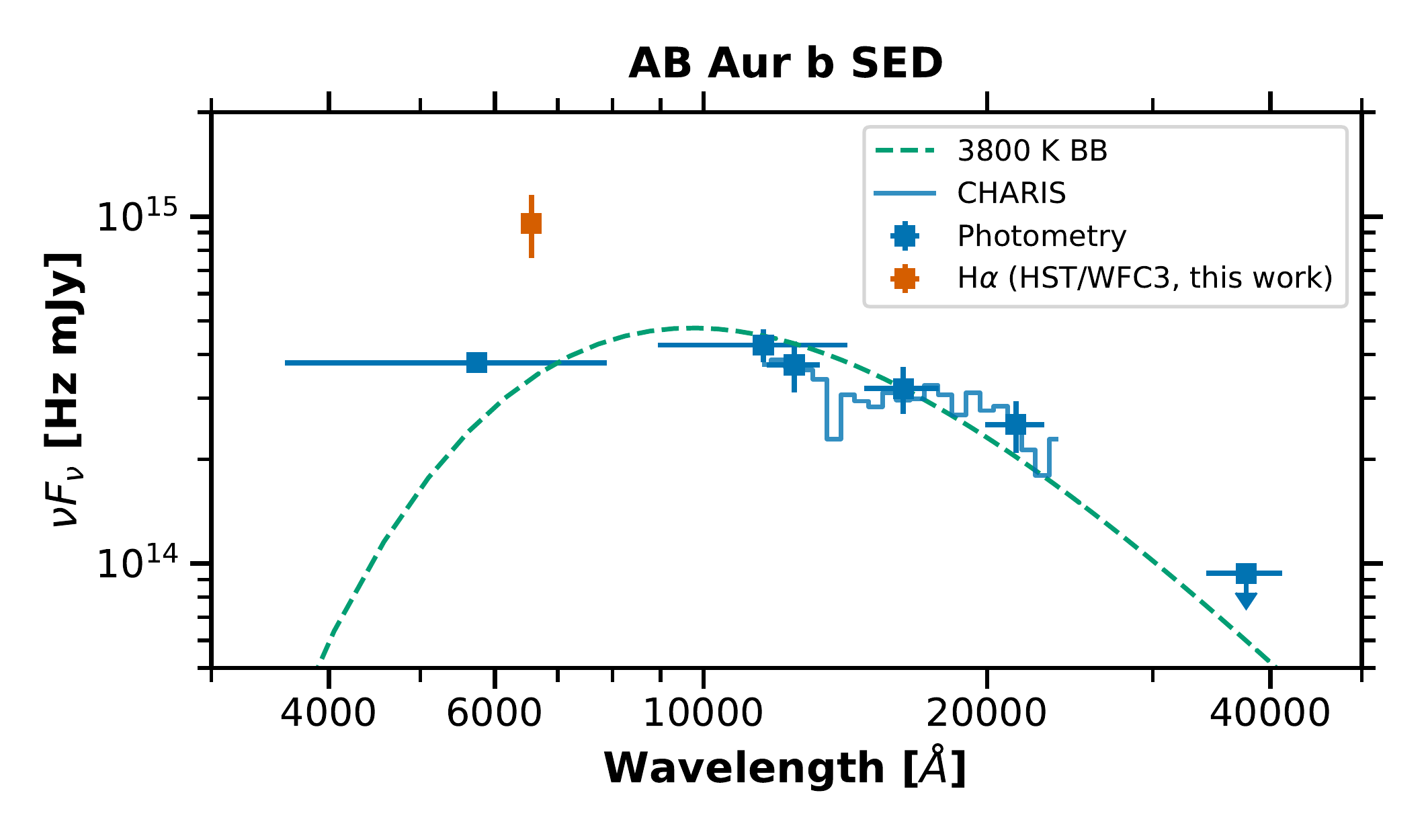}
    \includegraphics[width=\columnwidth]{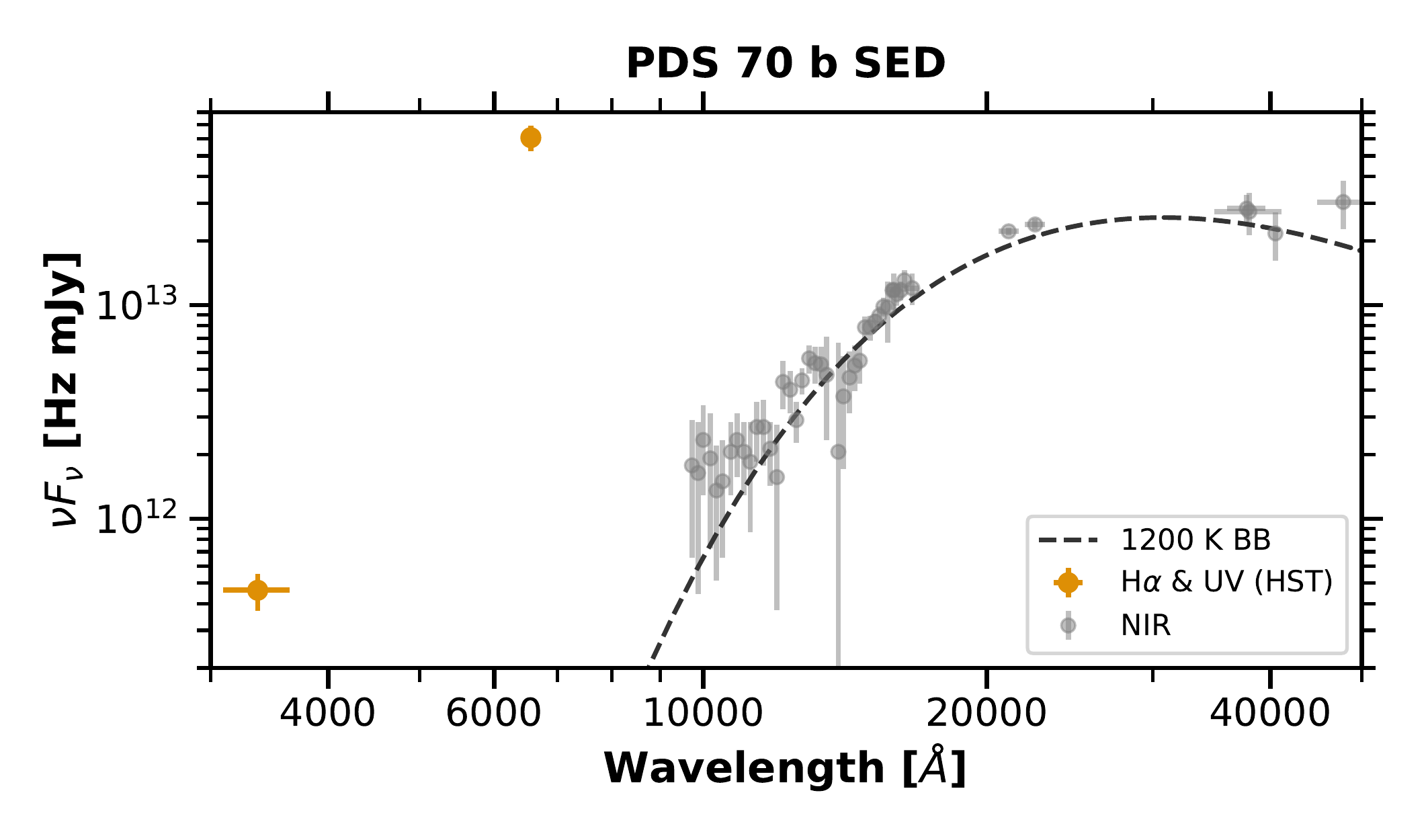}
    \includegraphics[width=\columnwidth]{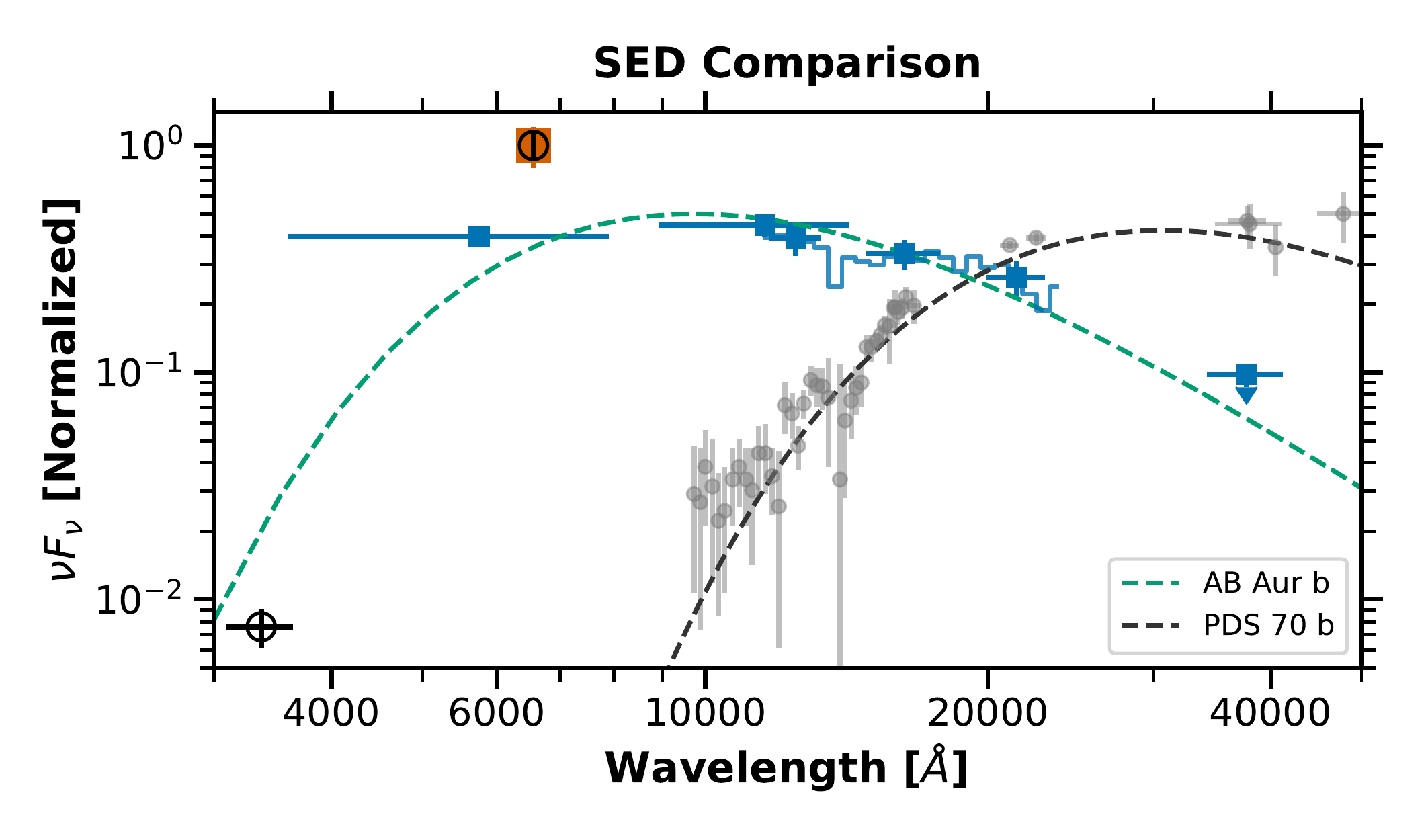}
    \includegraphics[width=\columnwidth]{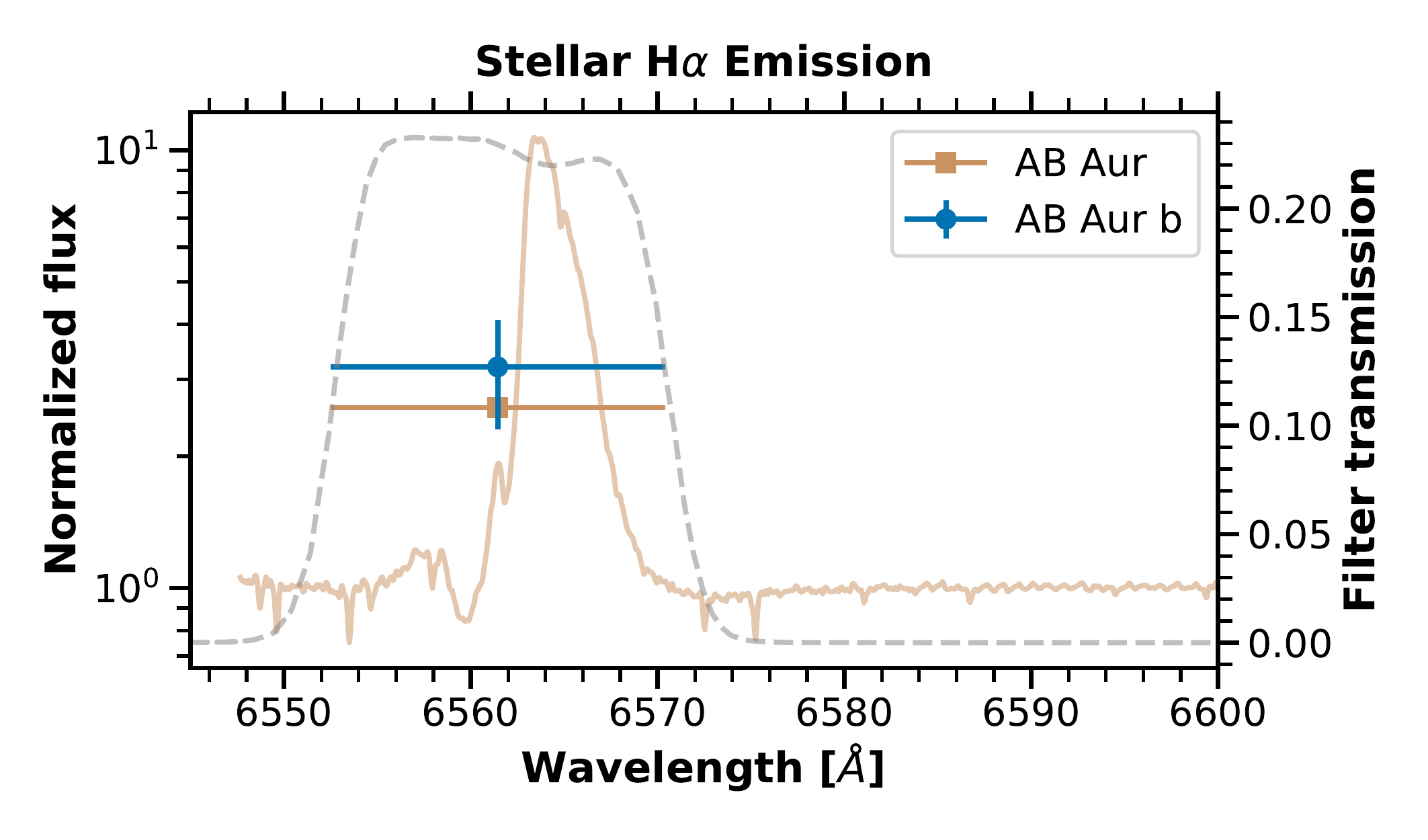}
    \caption{The SED of AB Aur b (upper left), PDS 70 b (upper right), a comparison of the two SEDs (lower left), and a comparison to the stellar \halpha{} emission of AB Aur (lower right). The SED of AB Aur b includes optical to NIR measurements from \citet[the blue squares and solid line]{Currie2022} and the H$\alpha$ flux from the combined HST data in this work (orange square). All AB Aur b fluxes are de-reddened based on $A_{V}=0.5$~mag and $R_{V}=3.1$. The best-fitting blackbody (3800~K) is shown in a green-dashed line. In contrast, the NIR SED of PDS~70~b (gray circles, data source: \citealt{Muller2018,Mesa2019a,Wang2020,Stolker2020}) is best-fit by a 1200~K blackbody (black dashed line). In the bottom-left panel, we normalize the two SEDs by their respective H$\alpha$ flux, so the \halpha points (orange square for AB Aur b, black circle for PDS 70 b) overlap. AB Aur b is bluer and has a stronger optical continuum. In the lower right panel, flux is normalized to the continuum (for AB Aur b, it is the HST/STIS flux). The \halpha of AB Aur b is consistent with the stellar value within $1\sigma$. }
    \label{fig:SED}
  \end{figure*}

  \section{Discussion}
  
  Rings and spirals in transition disks may introduce ambiguous protoplanet signals in high-contrast imaging observations. Emission from these structures may appear clumpy and even point-like after PSF-subtraction \citep{Follette2017,Rameau2017}. For example, the protoplanet candidates LkCa 15 bcd \citep{Kraus2012,Sallum2015} and HD 100546 bc \citep{Currie2015} were later found to be consistent with disk features (LKCa 15: \citealt{Currie2019}; HD 100546: \citealt{Follette2017} and \citealt{Rameau2017}). Because \halpha emission from an accreting planet distinguishes the planet from the disk in their spectral energy distributions (SED), \halpha images can provide an independent way of identifying protoplanets. For PDS 70 b and c, their \halpha flux densities are over 100 times that of the estimated optical continuum, which help confirm these objects as accreting protoplanets. In contrast, in the HD~100546 case, non-detections of \halpha-emitting point sources call into question the candidates as accreting protoplanets \citep{Follette2017,Rameau2017}. For AB Aur b, \citet{Currie2022} presented several lines of evidence to support the protoplanet identification: orbital motion has been detected, and the source's spectrum and polarimetric intensities differ from the rest of the disk. However, the companion resides in a region of extended scattered light and photospheric features appear to be absent in the SED \citep{Currie2022}. Below, we discuss whether the \halpha detection adds support to the protoplanet interpretation.

  The \halpha emission at the position of AB Aur b has been detected by Subaru/VAMPIRES \citep{Currie2022} and HST/WFC3 (this work). However, it is unclear whether the VAMPIRES detection is physically associated with AB~Aur~b due to uncertainties in the instrument's astrometric calibration \citep{Currie2022}. The astrometry of the WFC3 detection is well-calibrated and consistent with measurements in other bands. Therefore, we only adopt the WFC3 measurement for this discussion. The WFC3 band-averaged \halpha flux density of $f_{\nu}{=}1.5\pm0.4$~mJy is higher than the optical continuum measured by HST/STIS ($f_{\nu}{=}0.47\pm0.05$~mJy; $\lambda{=}0.57 \mu$m, $\Delta\lambda{=}0.43\,\mu$m) and Subaru/VAMPIRES ($f_{\nu}{=}1.13\pm0.37$~mJy; $\lambda{=}0.647 \mu$m), but the difference is moderate to marginal, respectively. The line-to-continuum flux ratio (defined as the band-averaged F656N flux divided by the continuum flux) is $3.2\pm0.9$ or $1.3\pm0.6$, depending on whether the STIS or the VAMPIRES flux is adopted for the continuum, much lower than the flux ratios ${>}100$ estimated for the PDS 70 planets, {as well as several accreting planetary mass companions \citep{Zhou2014,Eriksson2020,Stolker2021}}. AB Aur b's low line-to-continuum flux ratio is inconsistent with the accretion shock models that explains the observations of PDS 70 b and c \citep[e.g.,][]{Aoyama2018}, which produce pronounced \halpha lines.  As a result, the \halpha detection does not strengthen the interpretation that AB Aur b is an accreting protoplanet.
  
If the observed \halpha emission is indeed from an accretion shock, the shock also produces strong optical continuum emission that reduces the line-to-continuum flux ratio. Follow-up characterization of this source to measure ultraviolet and optical excess emission would provide useful comparisons to model SEDs and enable a diagnosis of the accretion shock interpretation \citep[e.g.,][]{Zhu2015}. Assuming that the observed F656N flux is entirely produced by the accretion shock of AB Aur b, we can determine the instantaneous mass accretion rate of the protoplanet based on the observed \halpha luminosity.  Adopting a line-of-sight extinction of $A_{V}=0.5$~mag (the same as the star, \citealt{Garcia-Lopez2006}) and assuming no additional extinction opacity, we find a \halpha line luminosity of $\Lhalpha=2.2\pm0.7\times10^{-5}\Lsol$. At this \Lhalpha, gas in the accretion flow should not lead to significant absorption \citep{Marleau2022}. The \Lhalpha value corresponds to a total accretion luminosity of $\log(\Lacc/\Lsol)=-2.8\pm0.3$ by assuming the planetary surface shock model \citep[][]{Aoyama2018,Aoyama2021} or $\log(\Lacc/\Lsol)=-3.5\pm0.3$ using the empirical relation for classic T Tauri stars \citep{Alcala2017}, respectively. Adopting a planetary mass of $9M_{\mathrm{Jup}}$ and radius of $2.7R_{\mathrm{Jup}}$ \citep{Currie2022}, these accretion luminosities can be translated to mass accretion rates of $\log(\dot{M}/(M_{\mathrm{Jup}}\mathrm{yr^{-1}}))$ between $-5.8$ and $-6.6$. These estimates roughly agree with the result $\log(\dot{M}/(M_{\mathrm{Jup}}\mathrm{yr^{-1}}))=-6.0$ obtained from SED fitting \citep{Currie2022}.

Alternatively, the observed \halpha flux can also be stellar light scattered by a compact disk structure or an envelope surrounding the protoplanet. In this case, the optical SED of the candidate protoplanet, which requires additional multi-band photometry to precisely determine, should mimic the SED of the star.  Remarkably, the line-to-continuum ratios of AB Aur (2.58) and AB Aur b ($3.2\pm0.9$, adopting HST/STIS photometry as the continuum) are consistent within $1\sigma$ (Figure~\ref{fig:SED}, bottom right panel). We note that this agreement might be serendipitous because AB Aur's \halpha emission is variable \citep{Harrington2007} and our comparison is based on asynchronous observations (the stellar spectrum is obtained 15 months earlier than the HST images). Our HST observations do not measure continuum flux and thus cannot simultaneously constrain the stellar line-to-continuum flux ratios.  Continuous and simultaneous \halpha monitoring of AB Aur and AB Aur b can help test this scattering scenario. \citet{Currie2022} concluded that the scattered light could not account for the infrared emission, because AB Aur b was undetected in infrared polarized images. If the \halpha flux is scattered light, the disk structure or dust envelope should predominantly scatter at short wavelengths such that scattered stellar emission is only significant in optical bands.

Finally, we emphasize the striking difference in the SED shape between AB Aur b and the PDS 70 planets. In Figure~\ref{fig:SED}, we compare the SEDs of AB Aur b and PDS 70 b and find that AB Aur b is significantly brighter and bluer. Fitting single blackbody models to the SEDs of AB Aur b and PDS 70 b (excluding the \halpha points) yield effective temperatures/radii of 3800~K/1.6~$R_{\mathrm{Jup}}$ and 1200~K/2.7~$R_{\mathrm{Jup}}$, respectively. These results correspond to over 1~dex difference in their bolometric luminosities ($\log(L_{\mathrm{bol}}/L_{\odot})=-2.3$ for AB Aur b \footnote{This value is higher than the one ($\log(L_{\mathrm{bol}}/L_{\odot})=-2.695\pm0.095$) reported by \citet{Currie2022}, because we adopted a blackbody rather than a photospheric model SED to calculate $L_{\mathrm{bol}}$.} and $\log(L_{\mathrm{bol}}/L_{\odot})=-3.8$ for PDS 70 b). We can compare the bolometric luminosity of AB Aur b to evolutionary track predictions while safely ignoring contributions from an accreting circumplanetary disk, because the expected disk luminosity is at least one order of magnitude lower\footnote{$\log(L_{\mathrm{CPD}}/L_{\odot})\approx-3.5$. This is estimated using Equation (5) of \citet{Zhu2015}, assuming $M=9M_{\mathrm{Jup}}$, $R_{\mathrm{in}}=7.5R_{\mathrm{Jup}}$ \citep{Currie2022}, and $\log(\dot{M}/(M_{\mathrm{Jup}}\mathrm{yr^{-1}})=-5.8$}. Assuming a hot-start model \citep[e.g.,][]{Burrows1997,Mordasini2017} and that AB Aur b has a planetary mass ($M{<}13M_{\mathrm{Jup}}$), AB Aur b has to be younger than 3~Myr to have such a high $L_{\mathrm{bol}}$. In contrast, $L_{\mathrm{bol}}$ of the PDS 70 planets are consistent with $1$ to $4$~$M_{\mathrm{Jup}}$ planets at an age of ${\sim}$5~Myr \citep{Wang2020,Stolker2020}. Because the isochronal age of AB Aur ($6.0^{+2.5}_{-1.0}$~Myr) is older than 3~Myr, the planetary interpretation of AB Aur b and the inferred young age might imply delayed planet formation in the AB Aur disk.

  \section{Summary}

  1. We observed the Herbig Ae/Be star AB Aur using HST/WFC3/UVIS in the F656N (\halpha) band in two epochs separated by 50 days. After subtracting the primary-star PSF, we detected a point-like source $0.6''$ away from the star in both epochs. In the most optimal reduction, the detection SNRs are 5.6, 4.4, and 7.8 in Epoch 1, Epoch 2, and the combined data-sets, respectively.

  2. The P.A. and separation of the companion are $182.5^{\circ}\pm1.4^{\circ}$  and $600\pm22$~mas relative to the host star. This location is consistent with the astrometry of AB Aur b, a recently reported protoplanet candidate. The candidate companions at separations of $2.75''$ and $3.72''$ are below our detection limits.

  3. The band-averaged \halpha flux densities are $0.9\pm0.3$, $1.2\pm0.5$, and $1.0\pm0.3$~$\times10^{-15}$erg/s/cm$^{2}$/$\angstrom$ ($F_{\lambda}$) or $1.3\pm0.4$, $1.8\pm0.7$, $1.5\pm0.4$~mJy ($F_{\nu}$) in Epoch 1, Epoch 2, and the combined images, respectively.

  4. In comparison to PDS 70 b and c, the \halpha-to-continuum flux ratio of AB Aur b is significantly lower, suggesting that an accretion shock that produces strong \halpha line that has flux density over 100 times the continuum is not visible from AB Aur b. Both planetary accretion and scattered stellar emission from either a compact disk structure or an envelope surrounding AB Aur b can contribute to the \halpha flux, and the origin of the \halpha emission remains unconstrained.

  5. Assuming that the observed \halpha emission is entirely powered by accretion onto the protoplanet and there is no gas or dust extinction from the disk nor from the accretion flow itself, we estimate an mass accretion rate of $\log(\dot{M}/M_{\mathrm{Jup}}\mathrm{yr^{-1}})=-5.8$ to $-6.6$, depending on the assumed $\Lhalpha-\Lacc$ relations. The accretion rate estimate suffers from significant systematic uncertainty due to the unknown origin of the \halpha emission and unconstrained extinction.


  The detection of \halpha emission from the candidate protoplanet AB~Aur b offers exciting opportunities to  understand the mass assembly processes of giant planets forming in a highly structured disk. These interpretations are conditioned on definitively resolving the origin of the \halpha emission. Disentangling the planetary accretion versus stellar scattering of \halpha emission possibilities could not only prove to be crucial in understanding the origin of AB Aur b, but also for the interpretation of future discoveries of protoplanets embedded in the disks of their host stars.


  \vspace{2em}
  We thank the referee for a constructive report that improves the rigor of the manuscript. We acknowledge excellent observing support from STScI staffs, especially J. Green and A. Armstrong. We thank Thayne Currie for the CHARIS spectrum and Gabriel-Dominique Marleau for helpful comments. Y.Z. acknowledges support from the Heising-Simons Foundation 51 Pegasi b Fellowship. B.P.B. acknowledges support from the National Science Foundation grant 450 AST-1909209, NASA Exoplanet Research Program grant 451 20-XRP20\_2-0119, and the Alfred P. Sloan Foundation. This research has made use of the NASA Exoplanet Archive, which is operated by the California Institute of Technology, under contract with the National Aeronautics and Space Administration under the Exoplanet Exploration Program.  The observations and data analysis works were supported by program HST-GO-16651. Supports  for  Program  numbers  HST-GO-16651 were provided by NASA through a grantfrom the Space Telescope Science Institute, which is operated by the Association of Universities for Research in Astronomy, Incorporated, under NASA contract NAS5-26555.

  \software{Astropy \citep{Robitaille2013}, Numpy \citep{Walt2011}, Scipy \citep{Virtanen2020}, Matplotlib \citep{Hunter2007}, pyKLIP \citep{Wang2015}}

\end{document}